\documentclass[review]{elsarticle}
\usepackage{lineno,hyperref}
\usepackage{graphicx}
\usepackage{epsfig}
\usepackage{epstopdf}
\usepackage{amsmath}

\journal{Journal of Physics in Medicine and Biology}

\begin{document}

\begin{frontmatter}

\title{Application of single walled carbon nanotubes for heating agent in photothermal therapy}

\author[mymainaddress]{Syahril Siregar \corref{mycorrespondingauthor}}
\cortext[mycorrespondingauthor]{Corresponding author}
\ead{siregar.syahril.p8@dc.tohoku.ac.jp}
\author[mymainaddress]{Israr Ul Haq}
\author[mymainaddress]{Ryo Nagaoka}
\author[mymainaddress]{Yoshifumi Saijo}
\address[mymainaddress]{Graduate School of Biomedical Engineering, Tohoku University, Sendai 980-8579, Japan }

\begin{abstract}
We present the theoretical investigation of the single walled carbon nanotubes (SWNTs) 
as the heating agent of photothermal therapy.  In our model, the SWNT is modeled by rigid 
tube surrounded by cancer cells. In this model, we neglect
the angle dependence of temperature and  assume that the length of SWNT 
is much longer than the radius of tube. We calculated the temperature rise of the SWNT 
and its surrounding cancer cells during the laser heating by solving 
one$-$dimensional heat conduction equation in steady state condition. We
found that the maximum temperature is located at the interface between SWNT
and cancer cells. This maximum temperature is proportional 
to the  square of SWNTs diameter and diameter of SWNTs depends 
on their chirality. These results extend 
our understanding of the temperature distribution in 
SWNT during the laser heating process and provide the suggested specification 
of SWNT for the improvement the photothermal therapy in the future.

\end{abstract}

\begin{keyword}
Photothermal therapy, single-walled carbon nanotubes, cancer, laser heating, heat conduction equation
\MSC[2016] 00-01\sep  99-00
\end{keyword}

\end{frontmatter}

\section{Introduction}
Over the past decade, the interaction between laser and nanoparticle has
received outstanding attention, because of the promised potential application 
for  the future cancer treatment \cite{avedisian2009nanoparticles,O'Neal2004171,Pissuwan200662,letfullin2006laser,pitsillides2003selective}. The cancer treatment using this
interaction is called the photothermal therapy (PTT).

PTT is recently investigated for cancer treatment, because of its safety 
as a consequence of using non$-$ionized radiation from laser. The PTT
process is given as follows: (i) the heating agent is injected into
the center of the cancer cell, (ii) the laser is irradiated directed 
to the cancer cells, (iii) the heating agent will absorb the 
laser energy, finally (iv) the temperature of heating agent and its surrounding will be
significantly increased. The rise in the temperature of cancer cells causes the
cancer cells to become hyperthermia and damaged. The cancer cells become 
damaged due to the destitute of blood supply, when 
its temperature reaches at least 41-47~$^0$C for several minutes \cite{huang2008plasmonic,svaasand1990physical}. 

The heating agent is very important in the PTT, because
irradiating the tumor cells without injecting the heating agent will not
damage the cancer cells \cite{moon2009vivo}. The heating 
agents commonly used in PTT are gold nanorod, and
gold nanosphere \cite{huang2008plasmonic, huang2006cancer, dickerson2008gold,
tong2009gold}. 

Even though, gold nanoparticles have been proved as potential heating 
agents of PTT from some previous works, gold nanoparticles have
limitation for heating agents of PTT. The absorption spectra of gold nanoparticles
lies in the  first near infra red (NIR) window (600 - 900 nm), furthermore, the 
penetration of light into heating agents in the first NIR window is less
efficient compared with the heating agents having absorption spectra in 
the second NIR window (1000 -1350 nm) \cite{tsai2013nanorod,smith2009second}.

Carbon based materials, such as graphene, graphene oxide and SWNTs  are
potential candidate for heating agents material of PTT, since carbon 
based materials have very strong absorption in the first and second
NIR windows \cite{moon2009vivo,yang2010graphene,doi:10.1117/1.3078803, robinson2011ultrasmall,yang2012influence}.

There are some merits of using SNWTs as heating agents of PTT, such as
the diameter of SWNTs is relatively smaller than diameter of most gold
nanoparticles, thus we can use SWNTs for cancer cells which located in the 
specific place. Moreover, the thermal conductivity of SWNT is ten times higher 
than gold nanoparticles. Therefore, based on its physical properties,
SWNTs is promised potential candidates for PTT heating agents.

Previous work about the experiment of PTT using SWNTs as heating agent, concluded 
that the SWNTs significantly enhanced the destruction of tumor cells 
without destructing the normal cells \cite{doi:10.1117/1.3078803}.

The theoretical simulations of PTT using laser and carbon nanotubes have been carried out
from previous work. Feng Gong
\textsl{et al.} investigated the rise in temperature of SWNTs during the 
laser heating process by using Monte Carlo approach \cite{gong2014mesoscopic}.
They made a cluster model, with cancer cells located in the center 
and SWNTs distributed in the surroundings of cancer cells, then investigated
the effect of SWNTs orientation distribution relative to the laser beam, such as
random, perpendicular, and parallel distribution \cite{gong2014mesoscopic}.  

Toshiyuki Nakamiya \textsl{et al.} investigated the thermal analysis
of multi-walled carbon nanotubes (MWCNTs) during pulsed laser heating by 
using finite element method \cite{nakamiya2008thermal}.   

In the present work, we made very simple model to calculate the 
rise in temperature of the heating agents and cancer cells by solving one$-$dimensional 
heat conduction equation within second order differential 
equation. Based on our model, we can provide 
the suggested characteristic of SWNT to improve the PTT for future cancer treatment.

\section{Properties of SWNT}

\begin{figure}[t]
\begin{center} 
\includegraphics[clip,width=0.67\textwidth]{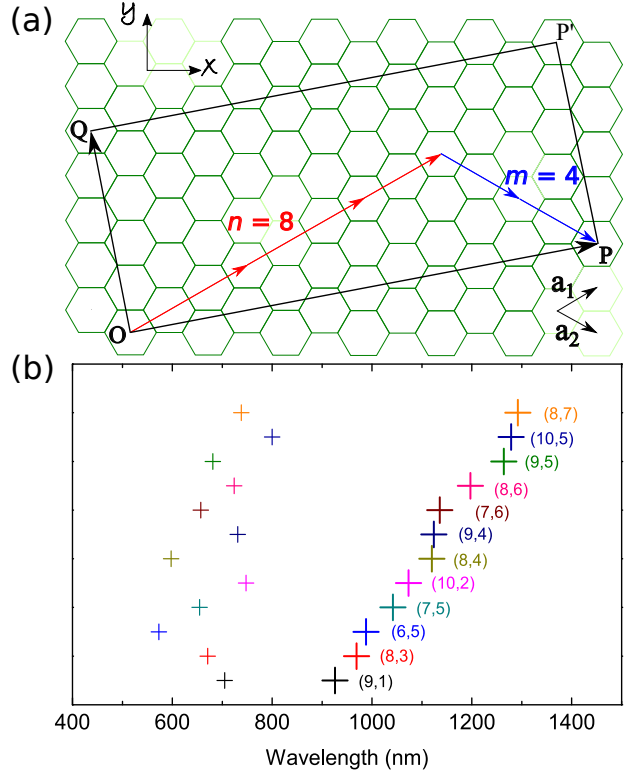} 
\caption{(Color online) (a) Geometry of (8,4) SWNT formed from graphene sheet with the graphene unit vectors $a_1\;$ and $\;a_2\;$.  If 
the $\mathrm{O}$ is joined to $\mathrm{P}$, and $\mathrm{Q}$ is joined to $\mathrm{Q'}$, the 
tube can be constructed.  (b) UV-vis-NIR spectrum of several chiralities of SWNT. There are two strong peaks with the stronger peak appear in the 900-1300~nm regions, data is adapted from reference \cite{tu2009dna}.}
\label{fig:fig_hexa}	
\end{center}
\end{figure}

The electronic and optical properties of SWNTs are depends on the their chirality. 
Chirality is described by chiral vector (n,m) which shows how the SWNT formed from 
the graphene sheet. In Fig.~\ref{fig:fig_hexa}~(a), the example of (8,4) SWNT 
formed from the graphene sheet is shown.

The electronic properties of SWNTs is classified into two groups, metallic and 
semiconductor. SWNTs can be classified as metallic if $\;n-m\;=3i$ and semiconductor if $\;n-m\;\neq 3i$,
with $i$ is integer \cite{hamada1992new,saito1992electronic}. Even so, in this work, we do not consider the electronics properties 
of SWNTs, based on the chirality we only take the information about diameter by using the 
function as shown in Tab. \ref{tab:tab1}.

In case of PTT, the information about the UV-vis-NIR spectra of heating agents is necessary, because
it demonstrates suggested wavelength of laser. In Fig.~\ref{fig:fig_hexa}~(b) we showed the UV-vis-NIR spectra adapted
from reference \cite{tu2009dna}. Generally, two peaks exist in UV-vis-NIR spectra for each chirality, the stronger peak
appear in the 900~nm~$-$~1300~nm (around the second NIR window) and the weaker peak appear in the 500~nm~$-$~800~nm (around the first NIR window). 

Although the thermal conductivity is sensitive to the temperature, in the present work, the thermal conductivity of SWNTs and cancer cells are assumed to be constant
to the temperature. The absorption coefficient of single molecules SWNTs was not measured nor calculated 
from previous work. Thus, we tune the absorption coefficient of single molecule SWNTs, so that 
the (6,5) SWNT gives the maximum temperature value around 57$^0$C, the similar temperature 
from experiment in the previous work \cite{doi:10.1117/1.3078803,antaris2013ultra},
and we use that value to calculate the temperature rise of another chirality.

\begin{table*}
\centering
\caption{Physical parameters of tissue, SWNT and laser.}
\label{tab:tab1}
\begin{tabular}{lll}
\hline
\multicolumn{3}{c}{Physical parameters}                                       \\ \hline
Thermal conductivity of tissue & $k_t$        & 0.567 $\mathrm{W/mK}\;$\cite{cooper1972probe} \\
Thermal conductivity of SWNTs  & $k_c$        & 3000$-$3500  $\mathrm{W/mK}\;$\cite{pop2006thermal} \\
Initial temperature            & $T_{\infty}$ & 37$^{o}$ C                   \\
Reflectivity                   & $R$          & 0.1                          \\
Absorption coefficient of SWNTs & $\alpha$     & 2.5 $\times 10^9 \;\mathrm{m^{-1}}$ \\
Laser intensity                & $I_0$        & $1 \times 10^{6} \;\mathrm{W/cm^2}$  \cite{qin2012thermophysical}            \\ 
Diameter of SWNT               & $d_{\mathrm{SWNT}}$ &$\left(n^2 + m^2 + nm \right)^{1/2}\;0.0783\;\mathrm{nm}\;$\cite{saito1998physical}  \\
Radius of SWNT							   & $a$          & $d_{\mathrm{SWNT}}/2$          \\  
The farthest considered distance &$b$          & $1000\;a$                       \\ \hline
\end{tabular}
\end{table*}

\section{Model}

\begin{figure}[t]
\begin{center}
\includegraphics[clip,width=0.6\textwidth]{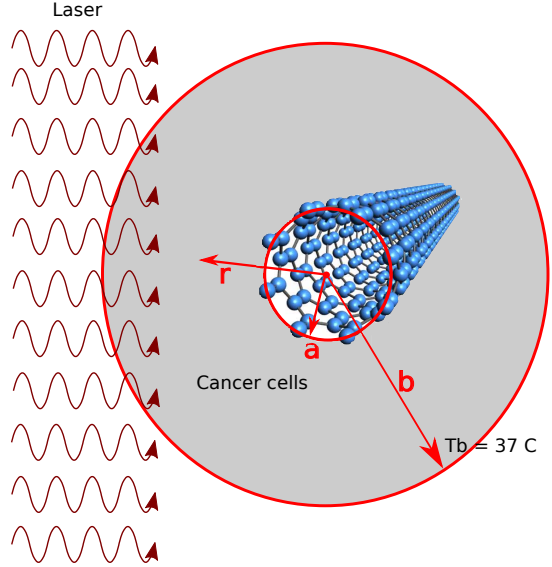} 
\caption{(Color online) Schematic model of SWNT and its surrounding (cancer cells) in PTT. The temperature in the outer cell with distance b from center is assumed to be temperature of body 37$^o$C. }
\label{fig:fig_model}	
\end{center}
\end{figure}

In this work, SWNT is modeled by rigid tube with radius $a$ surrounded 
by the cancer cells as shown in Fig. \ref{fig:fig_model}. In order 
to simplify the problem, we assume the length of tube is much longer 
than radius of tube, so we can neglect the dependence
on the length of tube. We also  neglect the angle dependence of temperature. Therefore,
the temperature rise during the laser heating only depends on radial distance. 

The PTT heating process is described by heat conduction equation given as 
\begin{eqnarray}
\rho_c c_c\; \frac{\partial T}{\partial t} &=& \frac{k_c}{r}  \frac{\partial}{\partial r} \left(r \frac{\partial T}{\partial r} \right) + q, \;\;\mathrm{for} \;\;  0 \leq r \leq a,  \label{eq:heat_conduction1}\\
\rho_t c_t\; \frac{\partial T}{\partial t} &=& \frac{k_t}{r}  \frac{\partial}{\partial r} \left(r \frac{\partial T}{\partial r}\right),  \;\;\;\;\;\;\;\mathrm{for} \;\;  r > a,
\label{eq:heat_conduction2}
\end{eqnarray}
where $\rho_{{c,t}}$ is the density of SWNT / cancer cells, $k_{c,t}$ is 
the thermal conductivity of SWNT/cancer cells, $T$ is temperature, and $q$ is the heat source
which represents the laser heating process. $r$ is radial distance measured from the center of the
tube. The heat source $q$ can be expressed as follow 
 
\begin{equation}
q = (1-R)I_0\; \alpha \exp(- \alpha z),
\label{eq:heating}
\end{equation}
where $I_0$ is the laser intensity, $\alpha$ is absorption coefficient of SWNT, $R$ is the reflectivity,  
and $z$ is the depth, measured from the interface to the center of the tube. 

We consider only steady state condition to simplify the problem, since many of interactions might be happen, if we consider 
the unsteady state condition, such as phonon-phonon, electron-phonon, and electron-heating 
interactions. The steady state condition of the heat conduction 
equation in eqs. (\ref{eq:heat_conduction1}) and (\ref{eq:heat_conduction2}), which are the time
independent equations given by, 

\begin{eqnarray}
k_c\;\frac{1}{r}\; \frac{d}{dr} \left(r \frac{dT}{dr}  \right) + q &=& 0,\;\;\;\;\;\;\mathrm{for} \;\;  0 \leq r \leq a, \label{eq:steady_state1} \\
k_t\; \frac{d}{dr}\; \left( r \frac{dT}{dr} \right) &=& 0, \;\;\;\;\;\;\mathrm{for} \;\;  r > a,
\label{eq:steady_state2}
\end{eqnarray}
with the boundary conditions
\begin{eqnarray}
\frac{d T}{d r} &=& 0, \;\; \mathrm{at}\; r = 0, \label{eq:b0} \\
T(a_{-}) &=& T(a_{+}), \label{eq:b1} \\
k_c \left(\frac{d T}{d r} \right)_{a_{-}} &=& k_t \left(\frac{d T}{d r} \right)_{a_{+}}, \label{eq:b2} \\  
T &=& T_{\infty}\;\;\; \mathrm{at}\;\;r \to \infty \label{eq:b3}.   
\end{eqnarray}
The temperature in the center of the tube should be finite, as described in eq. (\ref{eq:b0}).   
Eqs (\ref{eq:b1}) and (\ref{eq:b2}) explain that the temperature at 
the boundary should be continuous. The Farthest considered distance ($1000~a$) measured from the center 
of tube is assumed has the same temperature with the cells temperature 37~$^o$C ($T_{\infty}$)
and does not affected by laser heating, as can be seen in eq. (\ref{eq:b3}). 

\section{Analytical Results}
By doing some mathematical operations with considering the boundary condition in eqs. (\ref{eq:b0})$-$(\ref{eq:b3}), the solutions of eqs. (\ref{eq:steady_state1}) and  (\ref{eq:steady_state2}) 
can be obtained. The temperature as a function radial position is given by,

\begin{equation}
  T(r) = \left \{
  \begin{aligned}
    & \frac{q}{4 k_c} \left(a^2 - r^2 \right) + \frac{q a^2}{2 k_t} \ln \left( \frac{b}{a}\right) + T_{\infty} && \mathrm{for}\;  0 \leq r \leq a, \\
    & \frac{q a^2}{2 k_t} \ln \left( \frac{b}{r}  \right)  + T_{\infty} && \mathrm{for}\; r > a.
  \end{aligned} \right.
	\label{eq:analytic}
\end{equation} 

\begin{figure*}[ht] 
\begin{center}
\includegraphics[clip,width=1\textwidth]{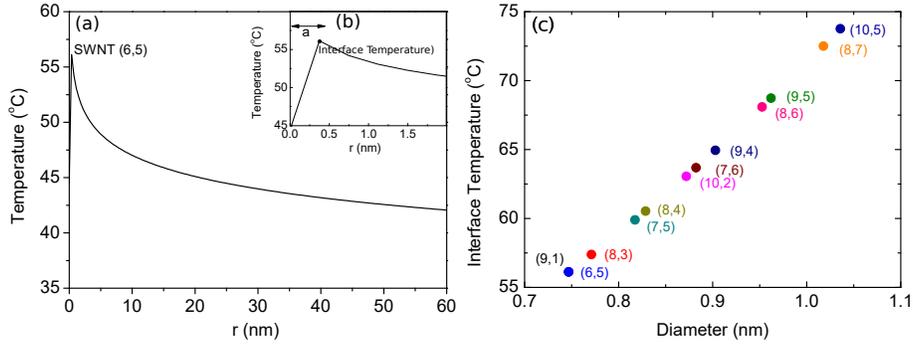} 
\caption{(Color online) (a) The calculated temperature rise as a function of radial distance measured from the center of (6,5) SWNT, (b) temperature in the interface between the (6,5) SWNT and the cancer cells, (c) the calculated interface temperature as a function of SWNT diameter for several chirallities.}
\label{fig:fig_result}	
\end{center}
\end{figure*}

The $q$ is appear even in the $r > a$, but the value of $z$ in eq. (\ref{eq:heating}) is zero for this region. The 
theoretical solution in eq. (\ref{eq:analytic}) shows that the temperature in the interface
is proportional to the square of SWNT radius. This result is in contrast with
previous work that concluded the smaller diameter have better heating efficiency \cite{gong2014mesoscopic}. In the 
present work, the larger diameter gives more depth for light to penetrate into the center of the tube, as 
a consequence, the tube will absorbs more energy from the light and convert it to the thermal energy. According 
to the plot in Fig.~(\ref{fig:fig_result})(b), the maximum temperature exist in the interface between tube and 
the cancer cells.

Based on theoretical result in eq.~(\ref{eq:analytic}), we plot the temperature as a function of 
radial distance for (6,5) SWNT, as shown in Fig.~\ref{fig:fig_result}~(a) and (b). We 
also plot the maximum temperature for several chiralities as shown in Fig.~\ref{fig:fig_result}~(c)
with selected chiralities following the previous work about the UV-vis-NIR spectra \cite{tu2009dna}.

\section{Conclusion}
Based on the theoretical calculation of laser heating process, we have shown that
SWNTs is potential candidates for PTT. The maximum temperature during the
laser heating process exist precisely in the interface between
SWNT and cancer cells. This maximum temperature is proportional to 
the diameter of SWNTs.  We believe that, our simple model can be used for 
further theoretical calculation of laser heating process of SWNTs 
by considering the transient effect.

\section*{Acknowledgement}
S.S. acknowledge the ImPACT project for
research assistantship.
I.U.H. is supported by MEXT scholarship.

\section*{References}

\bibliography{mybibfile}

\end{document}